\documentclass{article}

 \usepackage[preprint, nonatbib]{neurips_2025}
 \usepackage[style=apa]{biblatex}
 \newcommand{\citet}[1]{\textcite{#1}}
\newcommand{\citep}[1]{\parencite{#1}}
\usepackage[utf8]{inputenc} 
\usepackage[T1]{fontenc}    
\usepackage{hyperref}       
\usepackage{url}            
\usepackage{booktabs}       
\usepackage{amsfonts}       
\usepackage{nicefrac}       
\usepackage{microtype}      
\usepackage{xcolor}         
\usepackage{multirow}
\usepackage{listings}
\usepackage{tcolorbox}
\usepackage{float}
\tcbuselibrary{listings, breakable}
\usepackage{tabularx}
\usepackage{threeparttable}
\usepackage{caption}

\DeclareCaptionFormat{APAstyle}{
  \textbf{#1#2}\\[2pt] 
  \hspace{0.3em}\textit{#3}          
}

\newcommand{\fignote}[1]{%
  \vspace{2pt}%
  \par\noindent
  \parbox{\linewidth}{%
    \raggedright
    \textbf{Note.} #1
  }%
}

\title{Student Development Agent: Risk-free Simulation for Evaluating AIED Innovations}

\author{%
  Jianxiao Jiang \\
  School of Education\\
  Tsinghua University\\
  Beijing, China \\
  \texttt{jjx23@mails.tsinghua.edu.cn} \\
  \And
  Yu Zhang\thanks{Corresponding author. Address: 
417 Wennan Building, Tsinghua University, Beijing, China, 100084.
Email: zhangyu2011@tsinghua.edu.cn} \\
  School of Education\\
  Tsinghua University\\
  Beijing, China \\
  \texttt{zhangyu2011@tsinghua.edu.cn} \\
}

\begin{document}

\maketitle

\begin{abstract}
  In the age of AI-powered educational (AIED) innovation, evaluating the developmental consequences of novel designs before they are exposed to students has become both essential and challenging. Since such interventions may carry irreversible effects, it is critical to anticipate not only potential benefits but also possible harms. This study proposes a student development agent framework based on large language models (LLMs), designed to simulate how students with diverse characteristics may evolve under different educational settings without administering them to real students. By validating the approach through a case study on a multi-agent learning environment (MAIC), we demonstrate that the agent’s predictions align with real student outcomes in non-cognitive developments. The results suggest that LLM-based simulations hold promise for evaluating AIED innovations efficiently and ethically. Future directions include enhancing profile structures, incorporating fine-tuned or small task-specific models, validating effects of empirical findings, interpreting simulated data and optimizing evaluation methods.
\end{abstract}


\section{Introduction}

\subsection{Motivation}

Simulating student development in new learning environments has become an urgent necessity in the rapidly evolving landscape of AI-powered educational (AIED) applications. Educational innovations driven by Generative AI have not only emerged as a focal point in educational research but are also being increasingly implemented in practice, such as Khan Academy \citep{tassell2025mindfulness}, Massive AI-empowered Course (MAIC) \citep{zhang2025simulating}, and Alpha School \citep{dusseault2025ai}. These technologies and products aim to leverage technological advancements to facilitate learning and even transform education. However, empirical evidence regarding their effectiveness remains inconclusive. While some studies have attempted to validate the positive impacts of these technologies on students’ academic performance \citep{liu2025effects}, motivation \citep{renfeng2025motivational}, and higher-order thinking skills \citep{wang2025effect}, other research has highlighted potential risks, such as potential negative effects on students’ self-regulation \citep{darvishi2024impact} and autonomy \citep{dai2025students}, leaving a very complex and mixed picture.

Given the irreversible impact of education on student development, AI-driven educational innovations are inherently high-stakes. It is therefore imperative that all AIED innovations, including tutoring products, teaching methods, and learning approaches undergo rigorous controlled trials prior to widespread adoption, in order to assess their effects on both cognitive and non-cognitive outcomes, as well as to identify any potential harms. Nevertheless, due to ethical considerations and practical constraints, it is not always feasible to conduct long-term, large-scale experimental studies on new products and methods. Consequently, the challenge of rapidly and scientifically evaluating diverse AIED designs, as well as establishing robust assessment procedures, reasonable metrics and standards for these designs, has become a central research issue at the intersection of AI and education. It can provide an initial, preliminary evaluation of innovative products and models, safeguarding students’ normal development while also supporting the pace of AIED innovation \citep{xu2025classroom}.

Emerging generative AI techniques have enhanced the effectiveness of human simulation. Notable research include simulating users in human-computer interaction \citep{xiang2024simuser, wang2025effect}, and social simulations in social science \citep{park2023generative, park2024generative, zhang2025socioverse}. Although student simulation has already been employed in educational research and evaluation \citep{kaser2024simulated}, applying generative AI techniques to simulate students has become a trend \citep{xu2023leveraging}. Most existing studies in this area have focused on simulating student behavioral patterns within specific learning environments, or understanding and predicting students’ cognitive states, thereby enhancing the personalization of learning systems to improve students’ experiences and outcomes \citep{emond2023cognitive, xu2025classroom, zhao2023simulating}. Meanwhile, recent advances in self-evolving agent research offer methodological inspiration for educational simulations, suggesting new possibilities for modeling how students may evolve across learning experiences. Such techniques have been primarily applied in personal AI assistants and embodied agents for long-horizon tasks \citep{gao2025survey, fang2025comprehensive}. However, few studies have concentrated on simulating student development—namely, the evolution of student behaviors, achievements, or characteristics after novel designs, leaving a large research gap on how the AIED innovations impact student’s future development. 

In response to this gap, the present study proposes a student development agent design based on LLMs, to simulate the dynamic developmental trajectories of students under various educational settings. This agent is capable of integrating empirical findings acquired from real-world data with the generative capabilities of LLMs to rapidly and efficiently generate student changes. Such developmental simulation enables prospective evaluation of the potential effects of novel instructional applications or designs. This approach provides an evidence-based foundation for decision-making in technology facilitated learning and the optimization of AI in education.

\subsection{Related Work}

Emerging studies have leveraged LLMs to simulate students in educational contexts. For example, \citet{li2025exploring} incorporated demographic information, Big Five personality traits, goal commitment, motivation, self-efficacy, self-regulation, and several questionnaire responses into comprehensive student profiles, and developed a systematic pipeline for screening high-quality LLM-simulated students. The results demonstrated a group of simulated human-like profiles and behavior distributions. Similarly, \citet{xu2025classroom} designed a student simulation process based on a transferable iterative reflection module, and validated its effectiveness in modeling students’ online learning behaviors using real platform data. Similar designs have also been explored as summarized in Table~\ref{tab:simulation-review}. 

\begin{table}[h]
\centering
\caption{Summary of methods and key profile elements in recent LLM-based student simulation}
\label{tab:simulation-review}
\begin{tabularx}{\linewidth}{p{4cm}lX}
\toprule
Research (Authors)                              & Methods  & Key Elements in Profile                                                                                                       \\
\midrule
GPTeach \citep{markel2023gpteach}                   & Prompt   & context, student personas, recap,   message log                                                                               \\
Generative Students \citep{lu2024generative}       & Prompt   & Task, knowledge concepts master level                                                                                         \\
Eduagent \citep{xu2024eduagent}                      & Finetune & Student personas, gaze behaviors, motor   behaviors, cognitive states, test question performance, course/question   materials \\
TeachYou \citep{jin2024teach}                     & Prompt   & Knowledge state, reflection, response                                                                                         \\
HYP-MIX \citep{mannekote2025can}                & Prompt   & Persistence, Geometry Proficiency                                                                                             \\
Mathvc \citep{yue2024mathvc}                       & Prompt   & Name, gender, career, and Math skill   level                                                                                  \\
Personality-aware students \citep{liu2024personality} & Prompt   & Language abilities, the Big Five                                                                                              \\
Teachtune \citep{jin2025teachtune}                    & Prompt   & Knowledge component, goal commitment,   motivation, self-efficacy, academic stress                                            \\
Student Agent \citep{li2025exploring}                 & Prompt   & Demographics, psychological traits,   self-reported learning challenges, motivational factors                                 \\
ELL \citep{cai2025building}                          & Prompt   & Experience, skills, knowledge, actions              \\
\bottomrule
\end{tabularx}
\end{table}

While previous studies have demonstrated the potential of LLMs in simulating student behavior, most existing simulated student models remain static or only adapt within limited, predefined scenarios. They often lack the capacity to autonomously develop or evolve in response to novel instructional designs or long-term educational interventions—capabilities that are essential for modeling authentic human development. In addition, different student simulation studies defined various learning environments, inputs, and outcomes, which always focus on specific research interest and are inconsistent across different agent designs. This inconsistency undermines the comparability of current student agent models and highlights the need for a more generalized and unified design framework. Furthermore, most existing works primarily leverage the generative capabilities of LLMs and structured prompts, without sufficiently incorporating empirical findings from real-world educational data. Finally, current simulation studies tend to adopt context-specific evaluation methods, lacking a standardized evaluation procedure that is crucial for constructing reliable and convincing student development agents. 

\subsection{Research Questions}

Based on the reflections above, this study attempts to address the following research questions:

1)	What’s the framework of the dynamic profiles for student development agent design?

2)	How can empirical findings from real-world educational data be effectively acquired and utilized by the simulation of student development agents?

3)	Based on a case study, how accurately and reliably can LLM-based student development agents simulate student developmental trajectories and predict learning outcomes?

\section{Research Design}

\subsection{Design Framework of Student Development Agent}
\label{sec:des-fra}

\begin{figure}[]
  \caption{A Design Framework of Student Development Agent}
  \label{fig:ag-ds}
  \centering
  \includegraphics[width=.8\textwidth]{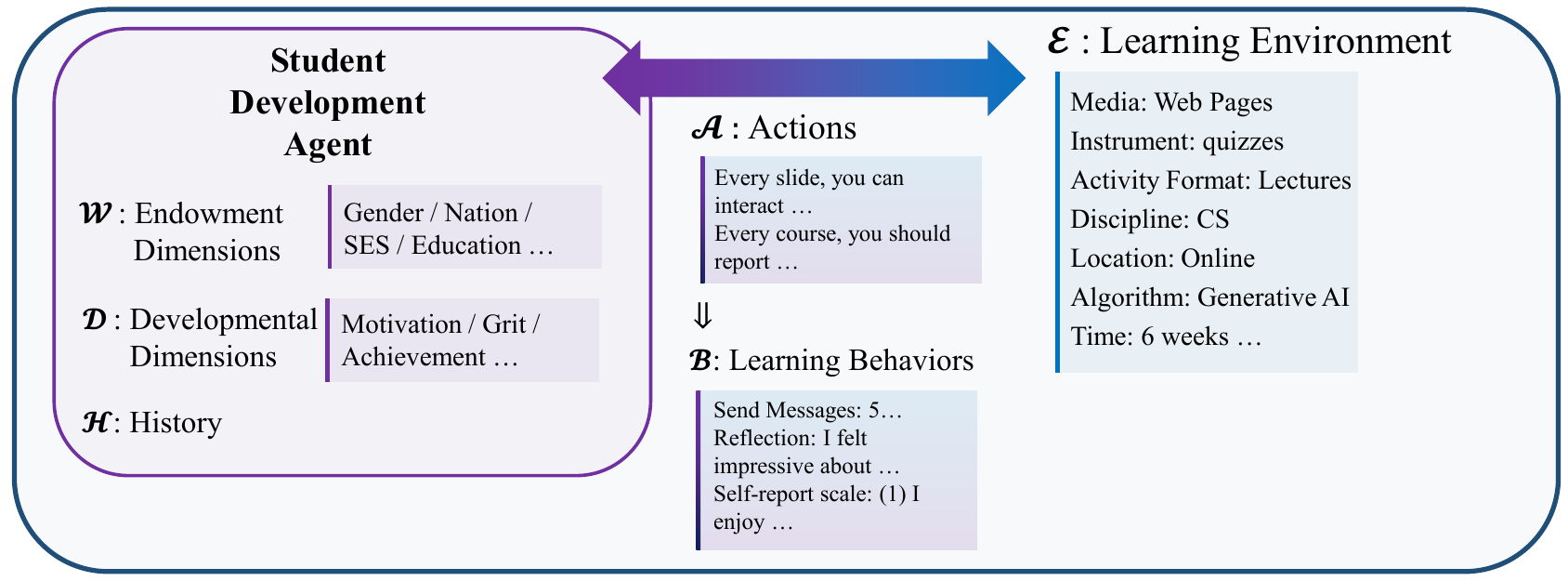}
\end{figure}

This study aims to propose a design framework for student development agent to model how students with diverse characteristics develop in respond to different learning environments. First, the \textbf{\textit{Input}} and \textbf{\textit{Output}} of the student agent are defined, following a conventional procedure in computer science \citep{luck2001conceptual}. Variables are denoted using mathematical symbols to clearly illustrate their interrelationships through mathematical equations. Inspired by educational theories and existing implementation of student simulation and self-evolving agents, the \textbf{\textit{Input}} to our student agent should include four key components including (1) the learning environment, (2) the endowment dimensions of students, (3) the developmental dimensions of students, (4) the actions defining how student agents can interact with the given learning environment. The \textbf{\textit{Output}} contains the simulated learning behaviors and changes in those developmental dimensions, encompassing both process-related and outcome-related measures. Furthermore, the history of the agent is also included in the design, containing past information of the agent along the simulation process. These variables are listed in Figure~\ref{fig:ag-ds}.

The first three key input components are grounded in classical educational theories and common practices. 

1)	\textbf{Learning environment} ($\mathcal{E}$) is defined as the physical, contextual, and cultural settings in which learning occurs, which commonly includes specific pedagogy, teaching strategy, teaching technology or learning platform embedded with clear learning strategies, student assessment methods, teacher quality, and etc..

2)	\textbf{Endowment dimensions} ($\mathcal{W}$) refer to key endowed features of students that are theoretically and empirically related to student learning and development. Commonly used characteristics include demographics (e.g., gender, socioeconomic status) and health conditions (physical or psychological conditions, such as disabilities, depression, autism, and etc.), which are usually stable and can categorize students into several subgroups, with relatively similar development trajectories within each subgroup.

3)	\textbf{Developmental dimensions} ($\mathcal{D}$) refer to measurable developmental changes in students that are crucial to educational purpose. These developmental dimensions are both outcomes for one period and inputs for the next period. Commonly measured outcomes include both cognitive (e.g., knowledge, cognitive skills, etc.) and non-cognitive outcomes (e.g., motivation, academic emotions, academic believes, personalities, etc.). 

The fourth component in the \textbf{\textit{Input}} is \textbf{Actions} ($\mathcal{A}$), which refers to how student agents interact with and respond to learning environments during the simulation processes. Researchers can define the timing and modality of agent actions, such as \textit{interacting with teachers and students in each class}, \textit{reflecting and reporting experiences after each class} or \textit{completing self-report questionnaires after the course}. These actions should be designed to occur within the learning environments, shaped by the instructional design and research objectives.

The two components in the \textbf{\textit{Output}} reflect the key focus of educational research and learning analytics.

\textbf{Learning behaviors} ($\mathcal{B}$) are student agents’ interactions and actions generated within the simulation processes, following the definitions given by Actions ($\mathcal{A}$). In other words, $\mathcal{B}$ is a collection of actual combinations (or sequences) of actions which are defined in $\mathcal{A}$ and are performed by student agents within the simulation.

\textbf{Results of developmental dimensions} ($\mathcal{D}$) describes the developmental dimensions attained by student agents as a result of the simulated learning behaviors. Notably, the structure of $\mathcal{D}$ follows the developmental dimensions defined in the Input, hence we use the same symbol to maintain consistency and show the self-evolving features. 

\textbf{History} ($\mathcal{H}$), as the last variable, refers to all the past information of the agent along the simulation process, as a common component in self-evolving agents’ design \citep{gao2025survey}. In our design framework, the History contains the past trajectories of the agent’s learning behaviors and developmental dimensions, as the learning environment and endowment dimensions are considered stable along the simulation process.

\subsection{Simulation Approach of Student Development Agent}

\begin{figure}[H]
  \caption{Simulation approach of student development agents}
  \label{fig:sim-ov}
  \centering
  \includegraphics[width=.8\textwidth]{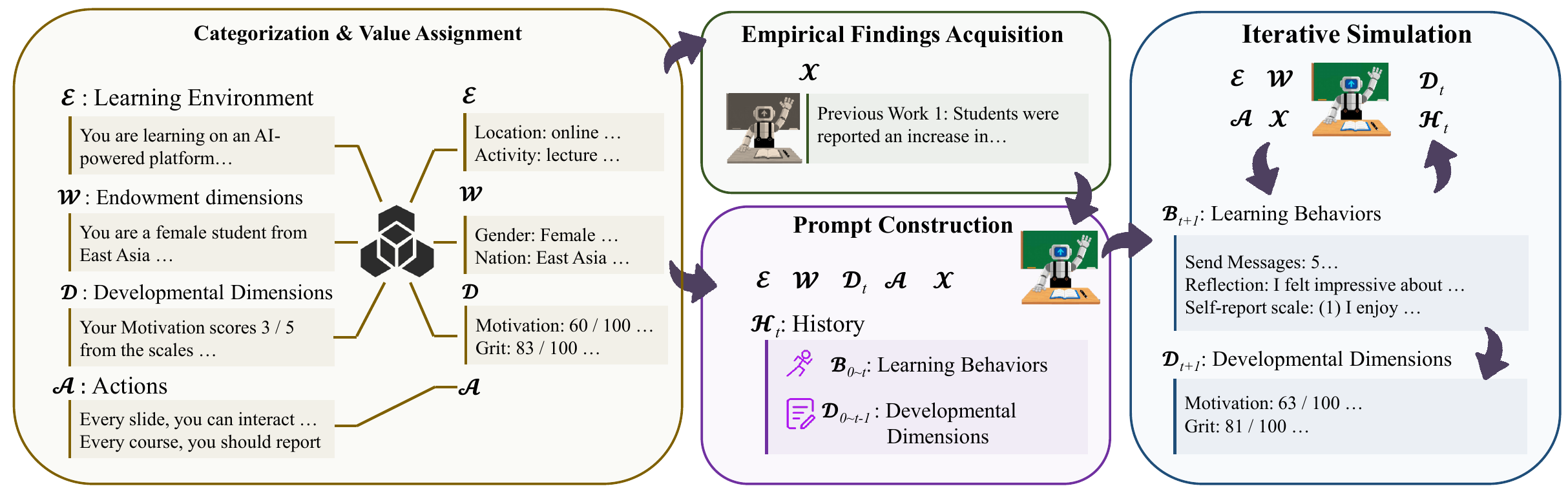}
\end{figure}

This section proposes an approach to leverage the student development agent for simulation, generating developmental results with a self-evolving mechanism. The simulation can be interpreted as:
\begin{equation}
    \{\mathcal{D}_{t+1}, \mathcal{B}_{t+1}\}= F(\mathcal{D}_{0\sim t}, \mathcal{B}_{0\sim t},\mathcal{E}, \mathcal{W}, \mathcal{A})
\end{equation}
, where subscript $t$ denotes the timepoint of $t^{th}$ period, and subscript $0\sim t$ denotes the time sequence from the initial period to the $t^{th}$ period, for each variable. \textbf{\textit{F}} denotes the simulation approach. To achieve the goal, the simulation approach includes four main parts: \textit{categorization \& value assignment}, \textit{empirical findings acquisition}, \textit{prompt construction}, and \textit{iterative simulation}. Figure~\ref{fig:sim-ov} shows an overview of the approach.

\subsubsection{Categorization \& Value Assignment}

\begin{figure}[H]
  \caption{Process of constructing general education categorization}
  \label{fig:ge-edu-cate}
  \centering
  \includegraphics[width=.9\textwidth]{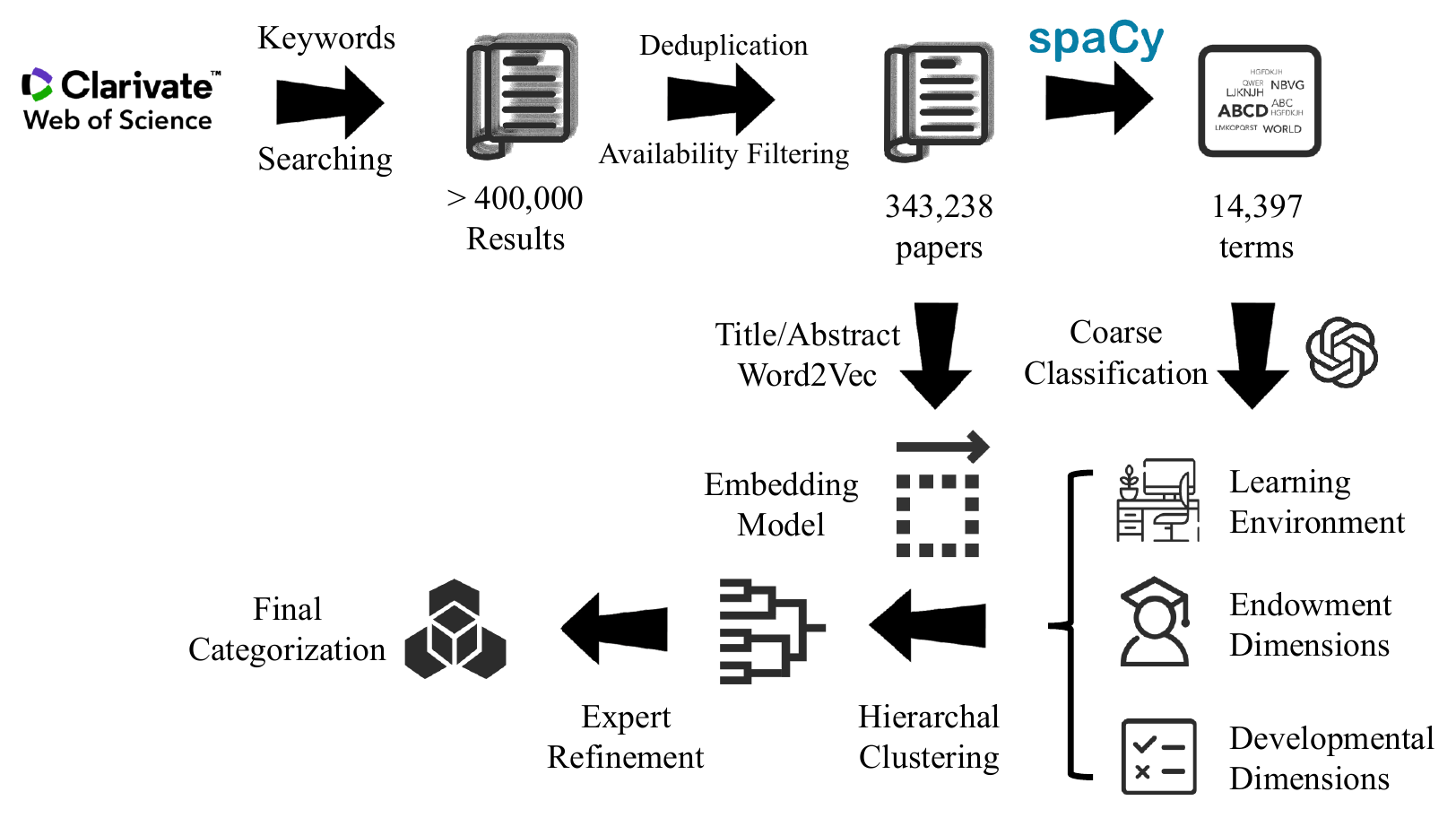}
\end{figure}

Over centuries of theoretical and empirical development, educational research has constructed a comprehensive body of knowledge encompassing diverse instructional designs, learning environments, technologies, and learner characteristics. With the rapid advancement of generative AI, AI-empowered technologies are expected to progressively extend into these varied educational contexts. Thus, rather than focusing on a review of current AIED applications, it becomes necessary to establish a complete and generalizable representation of educational environments and student features. Accordingly, we propose a general education categorization, which systematically organizes the dimensions of learning environments ($\mathcal{E}$), endowment ($\mathcal{W}$), and developmental aspects ($\mathcal{D}$) for the setup of student development agents. Leveraging over 300,000 research articles related to educational designs and empirical research collected from Web of Science, a pipeline including literature screening, term extraction, embedding generation, coarse categorization and semantic clustering was performed to construct the categorization structure. The resulting clusters were then reviewed and refined by two domain experts in education. Figure~\ref{fig:ge-edu-cate} displayed the complete process of the pipeline.

We conducted a literature retrieval and screening process using a comprehensive query in the Web of Science database, resulting in a large set of articles with titles and abstracts available. Key terms were extracted from the titles and abstracts and filtered based on frequency to build an initial vocabulary. These terms were then grouped into conceptual domains using LLMs, offering a coarse-grained categorization. Subsequently, word embedding techniques were applied to capture semantic similarities among the terms, and unsupervised clustering methods were used to identify and refine thematic clusters. Finally, the clusters were consolidated and labeled with the aid of domain expertise. Further methodological details are provided in the Appendix~\ref{apd:cate}.

Two experts in the field of education were then invited to evaluate the cluster results. An expert‐informed card-sorting evaluation was employed \citep{alam2021assessing}. From each cluster, the three most central terms (i.e., those closest to the cluster centroid) and the two most peripheral terms (i.e., those furthest from the centroid) were extracted. Domain experts were presented with these sampled terms on cards on a pre-established excel workbook. Experts were invited to sort the cards into categories based on their understanding of conceptual coherence. They were also encouraged to suggest adding or removing categories if the sampled items warranted such adjustments. After the process, Gwet’s AC1 was calculated to evaluate inter-expert reliability as new categories were allowed in the process which was not a suitable scenario for Cohen’s Kappa \citep{gwet2008computing}. Adjusted Rand Index (ARI) and Normalized Mutual Information (NMI) were calculated to evaluate external validity, quantifying agreement between experts and researcher \citep{hubert1985comparing, strehl2002cluster}. Finally, the experts’ feedback and suggested adjustments were integrated to produce the finalized general education categorization. Results were provided in Section~\ref{sec:res-cate}. Following this categorization, values are assigned to the student development agents accordingly.

\subsubsection{Empirical findings acquisition}
\label{sec:emp-fin-acq}

The empirical findings acquisition step is responsible for collecting useful data from real-world empirical research results, especially those with similar categories and variables with the Input of the agent to improve the performance of simulation \citep{wang2020generalizing}. Methods of finding relevant researches have been greatly enhanced by the emerging generative AI techniques \citep{fricke2018semantic, tong2024automating}. 

In this study, we propose two retrieval methods, taking advantage of the general education categorization. The first method is categorization-based keyword matching. Values assigned to the agent are considered as a set of keywords, which are then matched against the terms in the categorization to assess coherence with prior studies. Studies sharing the most relevant keywords are selected for further analysis. This structured approach enhances interpretability and ensures alignment with established educational contexts. 

Another method is using pre-trained LLMs (such as BERT, GPT-4o, etc.) to generate semantic embedding vectors for both the input query and the literature descriptions in the database. Cosine similarity is then computed between the input vector and all literature embeddings, and the records with the highest similarity in the semantic space are selected. This approach overcomes the limitations of structured coding when dealing with ambiguous or cross-domain terminology, thereby enhancing the flexibility and robustness of the acquisition method \citep{bhagdev2008hybrid}. 

The results from retrieval methods are denoted as $\mathcal{X}$ in our simulation process, and it depends on the Input of the design framework of student agents. Thus:
\begin{equation}
    \mathcal{X} = G(\mathcal{D}, \mathcal{E}, \mathcal{W}, \mathcal{A})
\end{equation}
, where \textbf{\textit{G}} denotes the method that searches the relevant empirical findings. 

\subsubsection{Prompt Construction}

Prompt techniques were utilized to complete the simulation processes. The prompts primarily consist of all variables defined in the Input (i.e., learning environments ($\mathcal{E}$), endowment dimensions ($\mathcal{W}$), developmental dimensions ($\mathcal{D}$) and actions ($\mathcal{A}$)). Additionally, empirical findings ($\mathcal{X}$) and history ($\mathcal{X}$) are appended. Empirical findings are acquired from real-world research data as described from Section~\ref{sec:emp-fin-acq}.

\paragraph{Learning Environments}

This component refers to the learning environment ($\mathcal{E}$) of the \textbf{\textit{Input}}, following the two-level categorization structure. Details in these subcategories would provide a clearer and more comprehensive picture for the agent, and improve the performance of the simulation.

\paragraph{Endowment Dimensions}

This component refers to the endowment dimensions ($\mathcal{W}$) of the \textbf{\textit{Input}}, also following the categorization to describe the endowed features of student agents that may not change through the simulation process. This component is similar to the profile settings in previous mentioned student simulation works.

\paragraph{Developmental Dimensions}

This component refers to the developmental dimensions ($\mathcal{D}$) of the \textbf{\textit{Input}}. In the agent simulation module, these are continuously updated as the simulation progresses. This dynamic updating mechanism enables the student agent to reflect changes in individual states during the simulation, thereby modeling students’ developmental trajectories in long-term interventions. 

\paragraph{Actions}

This component refers to the actions ($\mathcal{A}$) of the \textbf{\textit{Input}}. It uses natural languages to define the rules for agents to interact with the learning environments. Specifically, these actions can be entirely simulated or facilitated through existing tool interfaces. This component may clarify the timing, available tools and instructions, similar to LLM-based assistants using Model Context Protocol (MCP, \citep{hou2025model}).

\paragraph{Empirical findings}

This component incorporates the real-world educational research data ($\mathcal{X}$) retrieved in the section~\ref{sec:emp-fin-acq}, including detailed descriptions and quantitative results of these studies. The information is organized sequentially and embedded into the agent’s prompt in natural language. For numerical results, the items should be standardized to improve the comprehension of quantitative descriptions. 

\paragraph{History}

Similar to existing student simulation research and self-evolving agents, the agent simulation module in this study incorporates history mechanisms. The history component contains past developmental dimensions and generated learning behaviors. The history records the student’s historical behaviors and performance along the simulation process, providing continuity and contextual basis for the simulation. Necessary history compression methods are conducted if the length of the component exceeds the context window of the model. Thus, the history ($\mathcal{H}$) is generated by:
\begin{equation}
    \mathcal{H}_t = S(\mathcal{D}_{0\sim t-1}, \mathcal{B}_{0\sim t})
\end{equation}
, where \textbf{\textit{S}} denotes the possible compression or summarization process of the given long history. The subscript t denotes the time point of $t^{th}$ period, and subscript $0\sim t$ denotes the time sequence from the initial period to the $t^{th}$ period, for each variable.

\subsubsection{Iterative Simulation}

Based on the prompt construction described above, the student development agent is designed to exhibit dynamic and iterative developmental behaviors. The simulation process encompasses two steps:

\paragraph{Behavior Simulation}

According to the history of the agents ($\mathcal{H}$), the reference data ($\mathcal{X}$), and the given learning environment ($\mathcal{E}$), endowed dimensions ($\mathcal{X}$), and actions ($\mathcal{A}$), the student agents generated behaviors ($\mathcal{B}_{t+1}$) about how students perform under given conditions in the next time point. This process can be represented as:
\begin{equation}
    \mathcal{B}_{t+1}= L(\mathcal{H}_t, \mathcal{D}_t, \mathcal{X}, \mathcal{E}, \mathcal{W}, \mathcal{A})
\end{equation}
, where \textbf{\textit{L}} denotes the LLMs generation.

\paragraph{Developmental dimensions prediction}

Based on the simulated behaviors ($\mathcal{B}_{t+1}$), the student agents are instructed to report the updated states of developmental dimensions. This process can be represented as: 
\begin{equation}
    \mathcal{D}_{t+1}= L(\mathcal{B}_{t+1}, \mathcal{H}_t, \mathcal{D}_t, \mathcal{X}, \mathcal{E}, \mathcal{W}, \mathcal{A})
\end{equation}

Through these two steps, learning behaviors and developmental dimensions are updated accordingly. The updated states are then leveraged to update history, and assigned as the initial states of next iteration.

\subsection{Evaluation}

Evaluating the performance of simulation results is crucial for validating and optimizing student development agents \citep{kaser2024simulated}. We proposed three perspectives and corresponding metrics after synthesizing existing literature \citep{xu2024eduagent, li2025exploring}, with the first one being used in Section~\ref{sec:maicuse}.

\begin{itemize}
    \item Differences from empirical results. Differences between the outputs of the student agents and the real responses from human students are intuitive metrics to evaluate the performance of agent prediction. Quantitative metrics include Root Mean Square Error (RMSE), Mean Absolute Error (MAE), and paired statistical tests (t-test or Wilcoxon signed-rank test).

    \item Simulation authenticity. This metric focuses on whether student agents follow the static and dynamic profiles to generate responses. Interviews and questionnaires can be used to enable human experts to interact with student agents, so that experts can rate the authenticity to evaluate the simulation.

    \item Robustness. This metric focuses on whether student agents generate consistent predictions across multiple turns. Metrics include variances of outputs of repeating simulation processes, and also the variances of two metrics above across multiple simulation trials.
\end{itemize}

\section{Results of General Education Categorization}
\label{sec:res-cate}

The general educational categorization developed in this study comprises three main categories and 33 subcategories in total. The two-level categorization and terms set were both validated by domain experts.

\subsection{Categories in Learning Environment}

\begin{table}[H]
\centering
\caption{Categorization and examples of Learning Environment}
\label{tab:cate-e}
\begin{tabularx}{\linewidth}{lp{5cm}X}
\toprule
Subcategory           & Description                                          & Example terms                      \\
\midrule
Activity Format       & How learning process is   organized                  & Program, project, course, lecture  \\
Algorithm             & Algorithms used in the   learning systems            & Deep learning, machine learning    \\
Device                & Hardware students use for   learning                 & Blackboard, tablet, smart phone    \\
Discipline            & Disciplines that learning   materials belong to      & Algebra, architecture, electronics \\
Instrument            & how evaluation on students   is conducted            & Quiz, homework, assignment, task   \\
Location              & Physical or logical   locations where students learn & Classroom, library, online         \\
Media                 & How information is conveyed   to students            & Handout, workbook, textbook        \\
Mode \& Type          & The modes or types of education                      & Compulsory,   preparatory, formal  \\
Performance Metric    & Metrics evaluating the   environment                 & Inference speed, adaptiveness      \\
Service \& Support    & What environment provides   students with            & Teach, guide, scaffold             \\
Sociocultural Context & Sociocultural contexts where   students belong to    & Atmosphere, gender equity          \\
Task                  & What students do in the   environment                & Writing, speech, retell, read      \\
Technology            & Technologies used in the   environment               & Virtual reality, visualization     \\
Time                  & Description of the time of   the learning process    & Interval, frequency, short term   \\
\bottomrule
\end{tabularx}
\end{table}

This category encompasses the various physical settings, instructional structures, and cultural contexts in which students engage with educational processes. It includes fourteen subcategories, namely: activity format, algorithm, device, discipline, instrument, location, media, mode \& type, performance metric, service \& support, sociocultural context, task, technology and time. Detailed description and examples are listed in Table~\ref{tab:cate-e} following the alphabet order.

\subsection{Categories of Endowment Dimensions}

This category covers a range of student characteristics. It consists of 13 subcategories, namely: age, citizenship \& migration, educational stage, family, gender, language, physical disability, physical health, race \& ethnicity, region, religion \& culture, socioeconomic and talent. Detailed description and examples are listed in Table~\ref{tab:cate-w} following the alphabet order.

\begin{table}[H]
\centering
\caption{Categorization and examples of Endowment Dimensions}
\label{tab:cate-w}
\begin{tabularx}{\linewidth}{lp{5cm}X}
\toprule
Subcategory              & Description                                      & Example words                        \\
\midrule
Age                      & Age of students                                  & Adolescence, puberty, teenage        \\
Citizenship \& Migration & Nationality background and migration   status    & Foreign, multination, immigrant      \\
Educational Stage        & Educational level, stage,   phase and background & Undergraduate, senior, freshman      \\
Family                   & Family status of students                        & Adoptive, homeless, nonparental      \\
Gender                   & Gender and sex orientations                      & Male, female, bisexual               \\
Language                 & Language background                              & Bilingual, monolingual, multilingual \\
Physical Disability      & Physical disabilities that   students have       & Dyslexia, autism, deaf               \\
Physical Health          & Physical health status of   students             & Weight, appearance, body size        \\
Race \& Ethnicity        & Race and ethnicity of the   students             & Black, biracial, interracial         \\
Region                   & Where students live                              & Urban, suburban, rural               \\
Religion \& Culture      & Religious and cultural   background              & Religious affiliation, acculturation \\
Socioeconomic            & Socioeconomic status of   students               & Rich, poverty, middle class          \\
Talent                   & Special talents of students                      & Gifted, talented            \\
\bottomrule
\end{tabularx}
\end{table}

\subsection{Categories of Developmental Dimensions}

This category includes a diverse set of developmental features that span both cognitive and non-cognitive, physical and psychological dimensions. It contains 8 subcategories, namely: achievement, cognition, emotion, meta-cognition, motivation, physical health, social affective ability and traits. Detailed description and examples are listed in Table~\ref{tab:cate-d} following the alphabet order.

\begin{table}[H]
\centering
\caption{Categorization and examples of Developmental Dimensions}
\label{tab:cate-d}
\begin{tabularx}{\linewidth}{lp{5cm}X}
\toprule
Subcategory              & Description                                            & Example words                     \\
\midrule
Achievement              & Achievement and results of   learning                  & Achievement, proficiency, success \\
Cognition                & Cognitive skills                                       & Thinking, reasoning, memory       \\
Emotion                  & Emotional status of students                           & Stress, happiness, resilience     \\
Meta-Cognition           & Students' meta-cognitive   skills for self-development & Efficacy, self-regulation, agency \\
Motivation               & Students’ motivation for   learning                    & Interest, autonomy, intrinsic     \\
Physical Health          & Physical development of   students                     & Balanced diet, sleep quality      \\
Social Affective Ability & Social skills development                              & Collaboration, leadership         \\
Trait                    & Personal traits of students                            & Conscientiousness, neuroticism   \\
\bottomrule
\end{tabularx}
\end{table}

\subsection{Framework Validation}

To ensure the authority and reliability of the categorization, two experts in the field of education were invited to review and refine the results of clustering through a card-sorting process. The results were presented in Table~\ref{tab:res-cate}. Inter-expert consistency was high, with an overall Gwet’s AC1 of .96; specifically, values were .80 for Learning Environment, .98 for Endowment Dimensions, and .94 for Developmental Dimensions. After resolving the misunderstandings and conflicts between the two experts, the adjusted categorization results showed high agreement with the cluster results. The overall ARI and NMI were .97 and .95, respectively. Agreement by category was also strong: ARI values were .91 for Learning Environment, .96 for Endowment Dimensions, and .70 for Developmental Dimensions; NMI values were .96, .98, and .85, respectively. These findings further supported the reliability and accuracy of the categorization approach.

\begin{table}[H]
\centering
\caption{Results of inter-expert Gwet's AC1 and expert-cluster ARI and NMI indicators}
\label{tab:res-cate}
\begin{tabular}{lrrrrrr}
\toprule
\multirow{2}{*}{Main Category} & \multicolumn{2}{c}{Inter-expert} & \multicolumn{4}{c}{Expert-cluster}                      \\
\cmidrule{2-7}
                               & \multicolumn{2}{c}{Gwet’s AC1}   & \multicolumn{2}{c}{ARI}    & \multicolumn{2}{c}{NMI}    \\
                               \midrule
Learning Environment           & \multirow{3}{*}{.96}    & .80    & \multirow{3}{*}{.97} & .91 & \multirow{3}{*}{.95} & .96 \\
Endowment Dimensions           &                         & .98    &                      & .96 &                      & .98 \\
Developmental Dimensions       &                         & .94    &                      & .70 &                      & .85 \\
\bottomrule
\end{tabular}
\end{table}

In summary, the general education categorization was validated with strong inter-expert and clustering agreement, reaffirming the robustness of our approach. The next section shifts to a practical perspective, detailing an independent case study in a novel multi-agent classroom system to further investigate the feasibility of the student development agent framework.

\section{A Case Study: Predicting Students Development on a multi-agent learning environment (MAIC)}
\label{sec:maicuse}

To evaluate the practicality and effectiveness of the proposed student development agent, we conducted a case study within the MAIC environment—a novel online learning platform that allows students to interact with multiple teacher and student agents powered by LLMs \citep{yu2024mooc}. This case study serves as a minimal working example, designed to replicate findings from prior research while reporting the performance of our method in comparison to baseline approaches.

\subsection{Settings}

The MAIC system is an online learning platform featuring a set of LLM-driven agents that support teaching and learning through interactive, multi-agent dialogues \citep{yu2024mooc}. Human instructors collaborate with AI agents to design course materials, while students engage with six specialized agents—such as an AI Teacher, Sparker, Questioner, and Note Taker—each contributing unique instructional function. Director agents monitor classroom activity and adaptively manage agent participation based on student interactions and learning context. This setup fosters a dynamic, supportive environment where students receive diverse, personalized input and emotional encouragement during their learning process.

\subsection{Dataset}

The real-world student data used in this experiment were collected from a quantitative study conducted on the MAIC platform in June 2024 \citep{hao2025student}. The study involved 110 students who participated in a course titled \textit{Towards Artificial General Intelligence} on the MAIC platform. The dataset includes comprehensive log data and questionnaire data for all participants. The log data consist of students’ interaction histories on the platform, primarily focusing on their classroom discourse and participation. The questionnaire data include both pre- and post-course measurements across a wide range of dimensions, such as the Big Five personality traits, academic motivation, self-efficacy, grit, self-regulation, and technology acceptance.

To ensure representative coverage and validate the student development agents’ generalizability, 42 students were randomly selected from the full dataset. using a stratified sampling approach. Students were grouped according to a combination of non-cognitive skills (i.e., motivation, academic self-efficacy, grit, self-regulation, and technology acceptance measured by pre-course questionnaires), interaction behaviors (including total message counts and average message lengths), and pre-course test scores. From each stratum, one student was randomly selected to serve as a representative, thus maximizing diversity in both attitudinal and behavioral dimensions, in order to rigorously evaluate the predictive performance of the developed intelligent agent across a broad range of learner profiles.

\subsection{Methods}

\subsubsection{Student Profile Construction}

Based on the dataset described above, student profiles were constructed to include Big Five personality traits, academic motivation, self-efficacy, grit, self-regulation, and technology acceptance. Each attribute was standardized to a value between 0 and 100 according to the original questionnaire data.

\subsubsection{Interaction Simulation}

In the MAIC scenario, students can interact with intelligent agents through a browser interface. In this experiment, the simulation was implemented using the \textbf{\textit{Playwright}} Python library for browser automation. For each course slide, the agent's utterances—including the instructor's script and any subsequent interactions—along with the slide image, student profile, and platform usage instructions, were provided as prompts to the LLM to generate responses. Detailed prompt and examples are provided in the Appendix~\ref{apd:maicuse-prompt}.

\begin{figure}[H]
  \caption{Simulation process of student agents on the MAIC platform}
  \label{fig:maicuse}
  \centering
  \includegraphics[width=.9\textwidth]{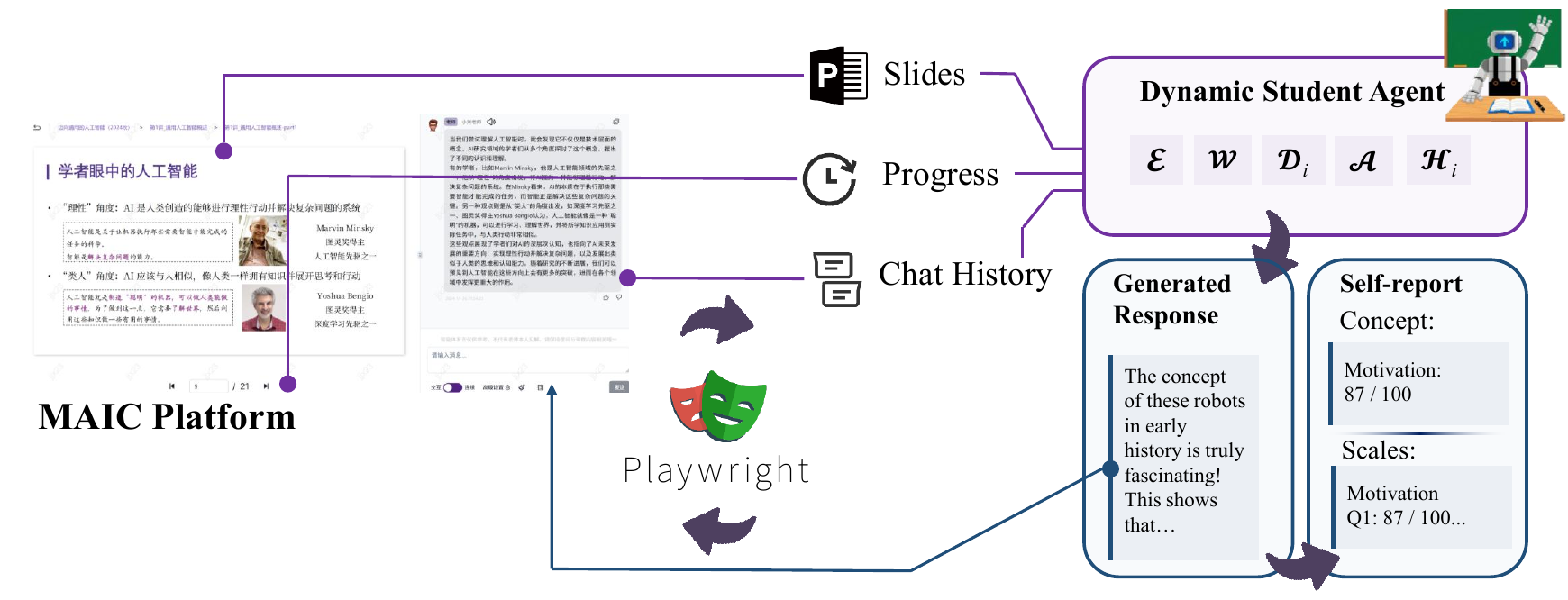}
\end{figure}

Considering that context length would grow rapidly in simulation and extend the number of max tokens that LLMs could accept, the simulation was limited to the first module. At the end of the module, the agent reflected on the interaction history and evaluated the post-course non-cognitive status. Two methods were adopted, including reporting values on the predefined concepts (\textbf{\textit{Concept}}) and filling up questionnaires of the scales (\textbf{\textit{scales}}). Figure~\ref{fig:maicuse} demonstrated the simulation process.

\subsubsection{Evaluation Metrics}

To evaluate the simulation performance of the student development agent, we compared the agent’s post-module questionnaire scores with those of the corresponding real students. As the student agent’s goal is to predict students’ development without administering the AIED innovation under investigation to real students, the baseline for comparison is the expected value—operationalized as the mean \textbf{pre-course} score for each dimension. Prediction error was evaluated using Root Mean Square Error (RMSE) and Mean Absolute Error (MAE), and paired t-tests were used to assess whether the prediction methods differed statistically. All scores were rescaled to 0-100. We also report a regression-based benchmark that requires administering the AIED innovation to real students and using their post-test outcomes to fit a multivariate linear regression; this is included only as a high-level reference.

\subsection{Results}

A total of 42 student agents was constructed in this experiment. The RMSE and MAE results comparing the baseline method (\textbf{\textit{mean}}) and the student development agent prediction (\textbf{\textit{concept}} and \textbf{\textit{scales}}) are reported in Tables~\ref{tab:maicuse-rmse} \& ~\ref{tab:maicuse-mae}. Besides, regression results (\textbf{\textit{regression}})that utilized post-test data are also reported for reference.

\begin{table}[H]
\caption{RMSE between prediction and true post-test values}
\label{tab:maicuse-rmse}
\centering
\begin{threeparttable}
\begin{tabular}{lrrrr}
\toprule
Dimension               & \multicolumn{3}{c}{\begin{tabular}[c]{@{}c@{}}Without   experiments\\ administered   on students\end{tabular}} & \begin{tabular}[c]{@{}c@{}}With   experiments\\ administered on students\end{tabular} \\
\cmidrule{2-5}
                        & \multicolumn{1}{c}{Baseline}                               & \multicolumn{2}{c}{Student Development Agent}                               & \multicolumn{1}{c}{Reference}                                                                                     \\
                        \cmidrule{3-4}
                        & \multicolumn{1}{c}{Mean}                                   & \multicolumn{1}{c}{Concept}                               & \multicolumn{1}{c}{Scales}                              & \multicolumn{1}{c}{Regression}                                                                                    \\
                        & \multicolumn{1}{c}{(1)}                                    & \multicolumn{1}{c}{(2)}                                   & \multicolumn{1}{c}{(3)}                                 & \multicolumn{1}{c}{(4)}                                                                                           \\
                        \midrule
Motivation              & 11.18                                  & \textbf{9.31}                                  & 17.41                               & 5.93                                                                                          \\
Academic Self-Efficacy  & 14.72                                  & \textbf{9.54}                                  & 10.83                               & 7.54                                                                                          \\
Grit                    & 13.30                                  & \textbf{10.74}                                 & 11.36                               & 7.85                                                                                          \\
Self-Regulated Learning & 8.92                                   & \textbf{8.47}                                  & 10.15                               & 5.37                                                                                          \\
Technology Acceptance   & 9.82                                   & 8.53                                  & \textbf{7.55}                                & 6.01                               \\
\bottomrule
\end{tabular}
\begin{tablenotes}
    \item \textbf{Note.} The best prediction results among methods \textbf{without} experiments administered on students are shown in \textbf{bold}.
\end{tablenotes}
\end{threeparttable}
\end{table}

\begin{table}[H]
\caption{MAE between prediction and true post-test values}
\label{tab:maicuse-mae}
\centering
\begin{threeparttable}
\begin{tabular}{lrrrr}
\toprule
Dimension               & \multicolumn{3}{c}{\begin{tabular}[c]{@{}c@{}}Without   experiments\\ administered   on students\end{tabular}} & \begin{tabular}[c]{@{}c@{}}With   experiments\\ administered on students\end{tabular} \\
\cmidrule{2-5}
                        & \multicolumn{1}{c}{Baseline}                               & \multicolumn{2}{c}{Student Development Agent}                               & \multicolumn{1}{c}{Reference}                                                                                     \\
                        \cmidrule{3-4}
                        & \multicolumn{1}{c}{Mean}                                   & \multicolumn{1}{c}{Concept}                               & \multicolumn{1}{c}{Scales}                              & \multicolumn{1}{c}{Regression}                                                                                    \\
                        & \multicolumn{1}{c}{(1)}                                    & \multicolumn{1}{c}{(2)}                                   & \multicolumn{1}{c}{(3)}                                 & \multicolumn{1}{c}{(4)}                                                                                           \\
                        \midrule
Motivation              & 9.12                          & \textbf{7.50}                         & 15.40                      & 4.71                                                                                         \\
Academic Self-Efficacy  & 11.36                         & \textbf{7.61}                         & 8.43                       & 6.40                                                                                         \\
Grit                    & 10.43                         & \textbf{8.19}                         & 9.10                       & 6.26                                                                                         \\
Self-Regulated Learning & 6.24                          & \textbf{6.59}                         & 8.22                       & 4.17                                                                                         \\
Technology Acceptance   & 7.52                          & 6.57                         & \textbf{6.07}                       & 4.81                                       \\
\bottomrule
\end{tabular}
\begin{tablenotes}
    \item \textbf{Note.} The best prediction results among methods \textbf{without} experiments administered on students are shown in \textbf{bold}.
\end{tablenotes}
\end{threeparttable}
\end{table}

Compared to the Mean baseline method, our approaches outperformed across all dimensions and both error estimation metrics. Specifically, the Concept method achieved the best metrics across all dimensions except Technology Acceptance, which was predicted the best using our Scales approach. However, when compared to the reference method Regression, our methods exhibited a performance gap. This is largely because the regression method incorporated students’ post-test scores, which naturally leads to stronger outcomes on the simulation indicators. As a reference method, it highlights the direction toward which our proposed approaches are striving. 

\begin{figure}[H]
  \caption{Distribution and paired t-test results between development prediction results from four methods and corresponding true values}
  \label{fig:maicuse-ttest}
  \centering
  \includegraphics[width=.7\textwidth]{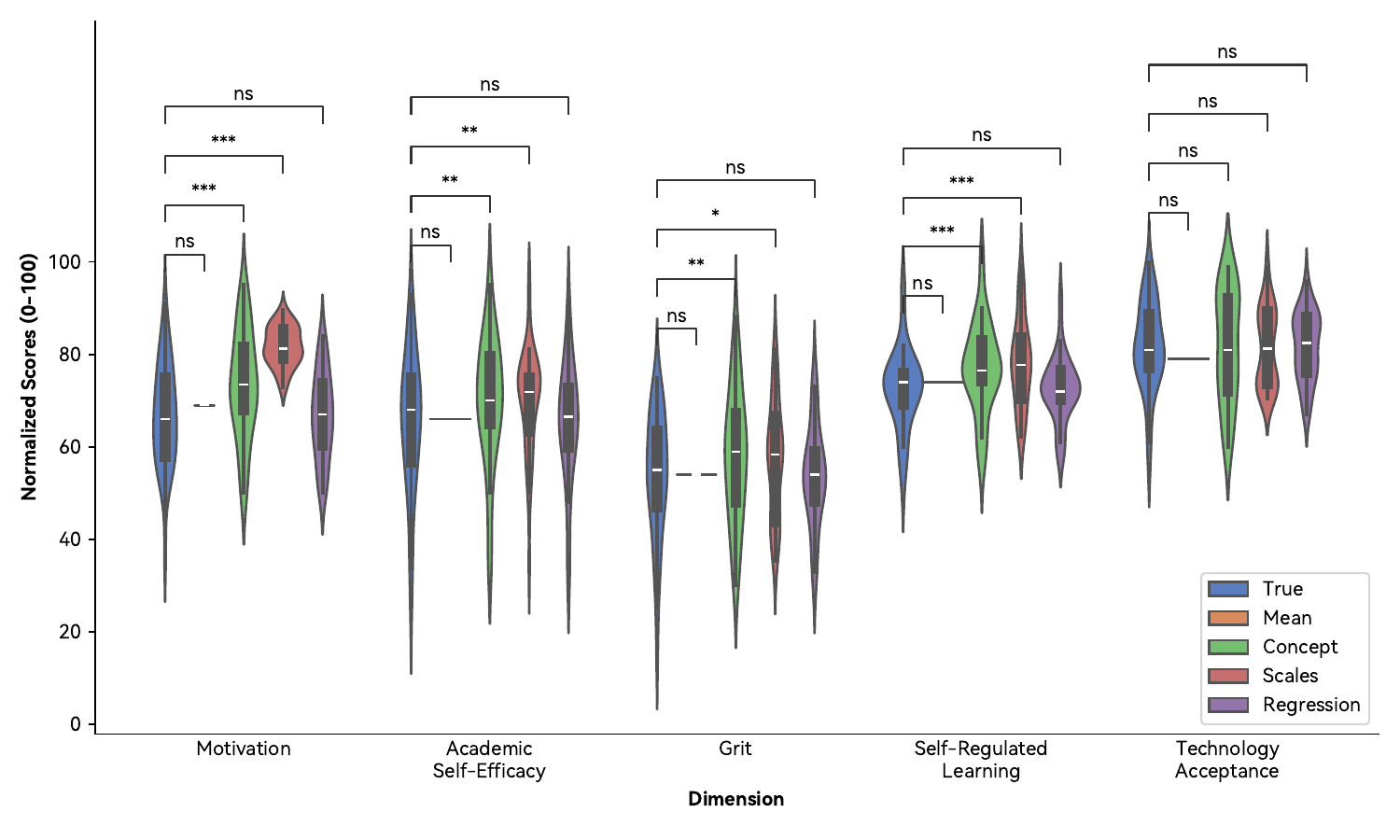}
  \fignote{ns: no significance, *: $p < .05$, **: $p < .01$, ***: $p < .001$.}
\end{figure}

Meanwhile, the distributions of post-test scores from real data and prediction results across the five dimensions are shown in Figure~\ref{fig:maicuse-ttest}. Among all dimensions, results of our student development agents showed no significance with true distribution only in Technology Acceptance dimension, which aligned with the results from RMSE and MAE estimations.

\section{Discussion}

\subsection{A step forward beyond student simulation: closing the loop}

The proposed development simulation framework marks a significant advancement beyond traditional student simulation research. Whereas most existing work focuses on generating behavior-level outputs, such as predicting student responses, classifying learning styles, or matching behavior distributions \citep{xu2024eduagent, li2025exploring}, our framework pivots to the long-term simulation of developmental outcomes and learning trajectories. Specifically, the student development agent executes actions (behaviors), observes the subsequent state transition (developmental outcome), and updates its internal state to inform the next round of actions. This closed-loop, agent-centric and iterative cycle allows the system to model long-period learning trajectories rather than merely simulating isolated behavior instances. By shifting the research focus from static behaviors to dynamic, state-dependent developmental outcomes, our framework lays the groundwork for more holistic and process-oriented educational simulations.

\subsection{A risk-free solution to safeguard real students}

Recent theoretical and empirical evidence highlighted the uncertain effects that AIED interventions could bring to real students \citep{baker2019education, dai2025students}. This uncertainty is critical: introducing untested systems into genuine learning environments—even for experimental purposes—creates a substantial risk of potential harm. This challenge raises significant ethical concerns regarding justice, student well-being, and responsible innovation.

Our proposed framework addresses this critical ethical gap by enabling a risk-free simulation process. This approach directly avoids exposing actual students to uncertain effects of new teaching approaches, assessment methods, or adaptive feedback mechanisms. Thus, the most sensitive stage of AIED innovation — initial prototyping and experimentation — can be conducted without risking real learners’ academic or psychological well-being. Furthermore, this methodology also aligns with recent calls in the literature for greater ethical scrutiny and “designed safety nets” in AIED research \citep{holmes2019artificial, luckin2016intelligence}. Our framework actively champions the core ethical goals of transparency, safety, and student protection, thereby advancing the broader mission of responsible and trustworthy innovation in education.

\subsection{Evolution brought to educational research}
The development of simulated students holds significant potential for advancing educational research. Traditional empirical studies in education often face challenges due to the high heterogeneity of both research subjects and learning environments. The diversity in intervention conditions, usage patterns, and student populations makes it difficult to produce generalizable and stable findings. Moreover, real-world studies are frequently constrained by factors such as high costs, long durations, and ethical concerns, all of which can undermine the validity and reliability of results \parencite[e.g.][]{chingos2011class}.

Our student development agent framework offers a novel solution by enabling the creation of standardized pipelines for testing and evaluating new educational settings. Similar to agent-based modeling platforms like NetLogo \citep{tisue2004netlogo} and TutorGym \citep{weitekamp2025tutorgym}, such pipelines can be implemented as a scalable and flexible online system. This platform would empower researchers to rapidly simulate a comprehensive spectrum of configurations, thereby accelerating experimentation and expanding the scope of analysis. Crucially, unlike human participants, simulated students are not subject to ethical limitations, permitting large-scale studies to be conducted with enhanced efficiency and responsibility.

In addition, the synthetic datasets generated by simulated students serve as a powerful complement to existing research methodologies such as meta-analysis and data mining \citep{zhan2023synthetic, khalil2025ai}. While meta-analyses provide crucial aggregated insights, they are often limited by effect size bias and critically lack sufficient raw data when examining emerging or underexplored educational contexts. Simulated students overcome this data scarcity by rapidly producing rich, high-fidelity synthetic data in these novel scenarios. For data mining and advanced analytical techniques, these synthetic datasets offer an unparalleled resource. They allow researchers to apply established analytical methods with rigor to investigate developmental patterns, refine predictive models, and even provide a direct basis to contrast the learning trajectories of human and artificial agents. This generated data effectively fills the gaps left by traditional real-world data collection.

\subsection{Future plans and directions}

\subsubsection{Optimizing the profile structure of student development agent}

The dynamic profile is a core component of the student development agent framework, providing the foundational context that shapes how the agent simulates behaviors and developmental changes. Our study proposed a fundamental structure consisting of the learning environment, endowment dimensions, developmental dimensions, actions and history, as well as empirical findings reference. This architecture is designed to capture the interplay of innate characteristics, environmental factors, and cumulative experience in the student's growth. However, the optimal structure and content richness of this profile remain open to further investigation. Existing simulation studies have demonstrated that rich, detailed profiles—sometimes constructed from extensive interview transcripts—can lead to highly realistic and context-sensitive simulations \citep{park2024generative}. Moreover, the history component presents a key area for optimization by leveraging recent emerging human-like memory techniques \citep{hou2024my, zhang2025survey}. These approaches offer a pathway to increase realism while proactively mitigating the risk of exceeding the LLM token limit. Determining whether these structural and memory optimization techniques yield comparable benefits within the student development agent framework is a critical question for future research efforts.

\subsubsection{Advanced simulation methods more than prompting}

While our current framework primarily relies on prompting strategies to simulate student development, a more fundamental advance lies in exploring model-level adaptations through techniques like finetuning or incorporating small models. Prompt engineering can guide behavior within a constrained context, but it has limited capacity to support iterative prediction, long-term development, and the emergence of stable agent traits \citep{junprung2023exploring, gu2023effectiveness}. Achieving a strong student development agent requires moving beyond surface-level prompting mechanisms.

Integrating finetuning techniques, particularly those based on reinforcement learning (e.g., reinforcement learning with human or synthetic feedback, RLHF), offers a promising path toward more robust developmental modeling \citep{ouyang2022training}. By exposing the model to sequences of simulated interactions and optimizing it based on outcome alignment or developmental plausibility, we can encode deeper behavioral patterns directly into the model's parameters \citep{xu2024eduagent}. This would enable agents to internalize long-term strategies, develop consistent personality traits, and respond adaptively to varying interventions, which is a prerequisite for achieving a truly developmental agent.

In addition, recent research in collaborative reasoning between LLM and smaller, specialized models introduces a complementary strategy \citep{patnaik2025learning}. Specialized small models can be trained or fine-tuned to maintain local memory, context tracking, or task-specific inference, while periodically querying a larger model for complex reasoning, abstraction, or behavioral generation. Such a hybrid approach not only improves computational efficiency but also supports a modular architecture where student agents can learn and evolve over time using a combination of lightweight context adaptation and powerful LLM-based generalization. This architectural synergy aligns perfectly with the core requirements of educational simulation, where sustained context, incremental growth, and adaptive behavioral change are essential for validity.

\subsubsection{Validation of empirical finding reference}

The empirical finding component was introduced in our profile structure, organizing empirical findings from similar previous educational research based on categorization values. This component effectively provides a behavioral and developmental reference for the agent under similar conditions, functionally mimicking a few-shot prompting mechanism. The effectiveness of few-shot learning has been widely validated across a range of general artificial intelligence applications \citep{wang2020generalizing}, lending initial support to this design choice. However, the application of empirical findings in the context of student simulation requires further validation, particularly given the high variability and context-dependence of developmental trajectories in authentic educational settings. While the goal of few-shot prompting is typically to align the model’s output with a given structure or expected form, this assumption may not hold in real-world learning scenarios. Because student development in authentic educational environments does not always conform to patterns reported in previous studies, over-reliance on empirical findings may constrain the model's capacity for simulating novel or unexpected behaviors. This, in turn, risks reducing its generalizability to new settings or its ability to respond creatively to unanticipated conditions. This challenge is analogous to the problem of overfitting in traditional machine learning, where a model becomes too closely tuned to its training data (or in this case, its few-shot examples) and fails to perform robustly on unseen or novel data \citep{ying2019overview, chamieh2024llms}.

Consequently, future research should carefully examine the balance between alignment and generalizability when integrating empirical findings in the agent design. While structured examples can improve stability and coherence in output, they must not override the model’s capacity to explore plausible but unobserved developmental paths. A nuanced validation strategy is needed to ensure that empirical findings support, rather than limit, the generative potential of LLM-based student agents. For example, systematically comparing simulation performance with and without the empirical findings component to isolate its causal impact, which is commonly adopted in machine learning to identify the contribution and necessity of specific model components \citep{sheikholeslami2019ablation}. Moreover, beyond the retrieval methods discussed in Section~\ref{sec:emp-fin-acq}, a systematic exploration of alternative data acquisition strategies is necessary to identify the optimal approach for populating the empirical findings reference.

\subsubsection{Trade-off of simulated data granularity}

A central challenge in employing LLMs to simulate student development lies in determining the appropriate granularity of the simulated data. Note that our case study benefited from the simplicity of the MAIC platform and the relatively narrow scope of student-platform interaction. Agents only need to accept the learning materials and chat messages displayed on the webpage, and generate text responses to complete the interaction with the platform. However, scaling the simulation to real classroom settings, broader school experiences, or feature-rich online platforms necessitates handling significantly more diverse and complex data types.

For example, in classroom learning scenarios, the simulation granularity can range from detailed modeling of students' thought processes in each lesson to more abstract reflections on a daily or weekly basis—especially in semester- or project-based simulations. While coarse-grained simulations may capture overarching developmental trajectories, they risk overlooking subtle yet critical learning behaviors and interaction patterns. Conversely, extremely fine-grained simulations, such as modeling every micro-decision within a classroom exchange, may introduce noise and computational complexity without proportionate theoretical gain. Balancing these levels of detail requires aligning the simulation design with specific research questions and the theoretical constructs under investigation. 

To systematically address this trade-off, future research should focus on developing a theoretically grounded data structure for simulated data.. This framework must establish standardized levels of simulated data granularity (e.g., event-level, session-level, daily-level) and detail the types of information richness (e.g., internal states, emotional responses, cognitive process traces) relevant at each level. Subsequent empirical studies can then methodologically investigate the effects of each granular level on simulation performance. Ultimately, this work is crucial for moving to a principled design methodology where the optimal granularity is defined not by technological constraint, but by the specific educational phenomenon being modeled. 

\subsubsection{Optimizing evaluation metrics and baselines}

Evaluating the outcomes of LLM-based student simulations requires not only robust statistical measures but also thoughtful methodological considerations. In this study, three fundamental aspects of metrics were proposed: differences from real data, simulation authenticity and robustness, and several indicators of differences from real data were validated in the provided case study in Section~\ref{sec:maicuse}.

Beyond the quantitative indicators employed in our presented case study, future work could extend evaluation by comparing the simulated learning trajectories with those of real students at the level of key behaviors and developmental dimensions. Such alignment would allow assessment of whether the simulated processes reproduce the critical mechanisms of learning rather than merely matching end-point outcomes. Another promising approach is to embed frequent but simple checks within the simulation process to ensure that the LLM’s behavior remains consistent with the traits and characteristics it claims to represent. 

A second consideration lies in the selection of baselines. While our case study adopted relatively simple baselines for comparison, these baselines themselves present limitations, as most existing prediction methods depend heavily on pre-test and process data. In scenarios where real-world data are sparse or unavailable, constructing reasonable baselines remains a significant challenge. One potential solution is to leverage findings from meta-analyses: by aggregating standardized effect sizes and parameters across large bodies of prior research, it becomes possible to generate theoretically grounded baselines that support more credible evaluations of LLM simulations in the absence of empirical data.

\section{Conclusion}

This study introduces a novel student development agent framework based on LLMs, aiming to simulate and predict how students with diverse profiles evolve under different learning environments. By integrating structured educational knowledge, empirical research data, and iterative simulation mechanisms, our approach advances beyond static prediction to model long-term developmental changes. The case study conducted on the MAIC platform validates the initial feasibility of this approach, showing competitive or superior predictive performance compared to traditional baselines. Moreover, the framework offers a scalable and ethical solution to evaluating AIED interventions without administering them to real students. Beyond technical feasibility, this work also contributes to methodological innovation in educational research. It demonstrates how language-based simulation can complement empirical studies, offering a standardized yet flexible tool for generating synthetic developmental data. The closed-loop design further enables exploration of causal mechanisms, adaptation strategies, and policy simulations.

Despite its contributions, the proposed framework represents only an initial step toward leveraging LLMs for simulating student development. Key challenges remain—particularly in designing more expressive and efficient dynamic profiles, generalizing the framework to more complex, real-world educational settings, and integrating fine-tuned models for long-term reasoning. Future research should explore richer memory architectures, reinforcement learning strategies, and collaborative reasoning mechanisms involving smaller, task-specialized agents. Building on the core design and the case study presented in this paper, the framework lays the groundwork for more holistic, adaptive, and scalable educational simulations—offering a promising foundation for research, intervention design, and intelligent tutoring in the era of generative AI.

{
\small

\printbibliography

}


\appendix

\section{Appendix}

\subsection{Detailed process of constructing general education categorization}
\label{apd:cate}

The approach to construct general education categorization consists of four steps: \textit{literature retrieval \& screening}, \textit{vocabulary extraction}, \textit{coarse classification} and \textit{semantic clustering}.

\subsubsection{Literature Retrieval \& Screening}

In this study, the Web of Science database was utilized as the primary source for literature retrieval. To ensure comprehensiveness and completeness, a refined search strategy was employed with the following query:

\begin{center}
\begin{tcolorbox}[colback=gray!10, colframe=black, boxrule=0.5pt, arc=1mm, breakable, width=.9\linewidth]
\begin{lstlisting}
(TS=(education OR instruction OR curriculum OR course OR student)) AND (TS=(support* OR scaffold* OR enhanc* OR improv* OR promot* OR facilitat* OR assist*)) AND (TS=(design OR environment OR framework OR concept OR platform)) AND (TS=(learn* OR performance OR "knowledge acquisition" OR understanding OR develop*)) AND DT=(Article OR Review)
\end{lstlisting}
\end{tcolorbox}
\end{center}

The initial search yielded over 400,000 articles. All records were exported and standardized into tables. These records were then screened according to usability relying on essential information, primarily the title and abstract. After these steps, 343,238 high-quality articles were retained as the foundation for subsequent analysis.

\subsubsection{Vocabulary Extraction}

For vocabulary extraction, the English abstract of each article in the initial dataset was selected as the analysis text. First, a tokenizer trained on large-scale English corpora from spaCy was used to ensure accurate recognition of domain-specific terminology. Then common stopwords (e.g., "the", "is", "of") and non-content words (such as numbers and symbols) were removed. Finally, a minimum frequency threshold was set (50 in our study) to eliminate noise and retain high-frequency keywords. This process resulted in an initial vocabulary set composed of 14,397 keywords.

\subsubsection{Coarse Classification}

To facilitate conceptual abstraction for framework construction, the initial vocabulary set obtained from the previous step was subjected to coarse classification using GPT-4o. The classification task was explicitly defined to assign each term to one of three major conceptual domains described in Problem Formulation: Learning Environment, Endowment Dimensions and Developmental Dimensions. Prompt templates were designed based on these predefined domains to guide the LLM in assigning each term to the appropriate category. Terms that did not fit any of the above domains or lacked sufficient contextual information were labeled as "Others". To improve processing efficiency, a batch processing approach was adopted, with each batch containing 50 terms. This iterative process continued until all terms in the vocabulary set were classified. Detailed prompt structures are as follows:

\begin{center}
\begin{tcolorbox}[colback=gray!10, colframe=black, boxrule=0.5pt, arc=1mm, breakable, width=.9\linewidth]
\begin{lstlisting}
You are an education research expert, please help me classify the research terms. Rules:
- Learning Environment (1): Refers to actively taken educational measures, such as teaching methods, technological tools, curriculum design, etc.
- Endowment Dimensions (2): Refers to characteristics of learners, such as age group, special groups, background features, etc.
- Developmental Dimensions (3): Refers to dimensions of the impact produced by educational interventions, such as academic performance (achievement), emotional expression (anxiety/happiness), non-cognitive abilities (motivation, efficacy, grit, regulation, acceptance, agency, autonomy, thinking), etc.
- Other (0): Clearly does not belong to the above categories or insufficient information.
Please follow the requirements below:
1. Evaluate each term separately without considering the relationship between terms.
2. Choose "Other(0)" when uncertain.
3. Strictly output in JSON format, as follows:
{{
    "categories": [
        {{
            "term": "term 1",
            "category": <1 or 2 or 3 or 0>
        }},
        {{
            "term": "term 2",
            "category": <1 or 2 or 3 or 0>
        }}
    ]
}}

List of terms to be categorized:
[50 terms]
\end{lstlisting}
\end{tcolorbox}
\end{center}

\subsubsection{Semantic Clustering}

The abstracts extracted from the aforementioned literature were used as the training corpus for the Word2Vec model. The Skip-gram algorithm was employed to construct word embeddings, with the following parameter settings: a window size of 10, a minimum word frequency of 5, and an embedding dimension of 300. All terms from the three coarse-grained categories were input into the trained Word2Vec model to obtain their corresponding vector representations. For clustering, agglomerative hierarchical clustering was applied separately to each of the three categories. Cosine distance was used as the metric for measuring similarity between word vectors, and the "Average" linkage method was selected as the aggregation criterion. Rather than relying on metrics such as silhouette scores or the elbow method to determine the optimal number of clusters, we adopted a threshold-based approach tailored to each clustering task. This strategy balances the trade-off between cluster quality and quantity, thereby enhancing the interpretability of results and supporting efficient manual review and refinement by researchers or domain experts.

\subsection{Example Prompts of Student Development Agent in MAIC case study}
\label{apd:maicuse-prompt}

The prompts used in the case study consists of two parts: the system prompt and user prompt. The system prompt is responsible for providing information on variables defined in Section~\ref{sec:des-fra}, and the user prompt is designed to instruct the agent to generate behaviors or complete self-reports.

\subsubsection{Example system prompt}

\begin{center}
\begin{tcolorbox}[colback=gray!10, colframe=black, boxrule=0.5pt, arc=1mm, breakable, width=.9\linewidth]
\begin{lstlisting}
You are a student participating in an online course. Your task is to learn the course content by interacting with the course website. You will be provided with platform instructions, your profile, and the chat history. You will also be provided with the current slide of the course. Your goal is to understand the content of the slide and engage in discussions with other students and instructors, according to your profile. You should respond with text messages you want to say confronting the current slide and the chat history.

# Your Profile
Name: [name]
Role: Learner
Course: Towards General Artificial Intelligence
## Your Non-Cognitive Skills
Motivation: [motivation]
Academic Self-Efficacy: [academic self-efficacy]
Grit: [grit]
Self-Regulated Learning: [self-regulated learning]
Technology Acceptance: [technology acceptance]
## Your Personal Traits
Neuroticism: [neuroticism]
Conscientiousness: [conscientiousness]
Agreeableness: [agreeableness]
Openness: [openness]
Extraversion: [extraversion]

# Platform Instruction
The MAIC platform is an online learning environment designed to facilitate the learning process through interactive elements. The environment contains slides on the left side, a chat area on the right side. On default you are in interactive mode. In this mode, the chat area allows you to interact with many students and instructors, you can ask questions, share insights, and engage in discussions. The teacher will first explain the current slide, and maybe there are other students who will propose questions or comments. You can also ask questions or share your insights in the chat area. The teacher or other students may then respond to your questions and provide feedback on your insights. If you do not want to interact with the teacher or other students, you can simply say "continue" to move on to the next slide.

Please respond following these rules:
1. Because you are a student, you should act according to your profile.
2. In-class messages should not be too long, and should be concise. Generally, each message should contain about 3 sentences.
3. Do not contain much descriptive information related to the slide or knowledge, provide key insights or proposals.
4. Do not repeat the content of the slide or the chat history, say "continue" if there is nothing new to say.
5. If you do not want to interact with the teacher or other students, you can simply say "continue" to move on to the next slide.

# Chat History
The chat history is a record of your interactions with the platform during current slide. It includes messages from you, the teacher, and other students. You can refer to the chat history to recall previous discussions and insights shared during the course.
[history]
\end{lstlisting}
\end{tcolorbox}
\end{center}

\subsubsection{Example user prompt}

\paragraph{Behavior simulation}

\begin{center}
\begin{tcolorbox}[colback=gray!10, colframe=black, boxrule=0.5pt, arc=1mm, breakable, width=.9\linewidth]
\begin{lstlisting}
According to your profile, the platform instructions, the chat history and current slide, please provide your response in Chinese.
# Your Response (In Chinese Please)

\end{lstlisting}
\end{tcolorbox}
\end{center}

\paragraph{Developmental dimensions prediction}

For the \textbf{\textit{concept}} method:

\begin{center}
\begin{tcolorbox}[colback=gray!10, colframe=black, boxrule=0.5pt, arc=1mm, breakable, width=.9\linewidth]
\begin{lstlisting}
# Your Task
According to your background and the chat history, please reflect on the lesson you just experienced with the platform, and then:
1. Reflect on your learning experience in this lesson. ** You should focus on your learning experience, specifically how you interacted with the platform and the agents you met. **
2. Provide your current status after this lesson, including:
- motivation
- academic self-efficacy
- grit
- self-regulated learning
- technology acceptance
All values should be between 0 and 100.

# Response Format
Your response should be in the following format:
{{
    "reflection": "Your reflection on the learning experience",
    "status": {{
        "motivation": 0 to 100,
        "academic_self_efficacy": 0 to 100,
        "grit": 0 to 100,
        "self_regulated_learning": 0 to 100,
        "technology_acceptance": 0 to 100
    }}
}}

\end{lstlisting}
\end{tcolorbox}
\end{center}

For the \textbf{\textit{scales}} method:

\begin{center}
\begin{tcolorbox}[colback=gray!10, colframe=black, boxrule=0.5pt, arc=1mm, breakable, width=.9\linewidth]
\begin{lstlisting}
# Your Task
According to your background and the chat history, please reflect on the lesson you just experienced with the platform, and then:
1. Reflect on your learning experience in this lesson. ** You should focus on your learning experience, specifically how you interacted with the platform and the agents you met. **
2. please conduct the post-test for [target dimension] after the course.
[target dimension description]

# Response Format
Your response should be in the following format:
{{
    "reflection": "Your reflection on the learning experience",
    "scale": [target dimension scales]
}}

\end{lstlisting}
\end{tcolorbox}
\end{center}


\end{document}